# Reputation Systems: An Axiomatic Approach

**Moshe Tennenholtz**
Faculty of Industrial Engineering and Management
Technion–Israel Institute of Technology
Haifa 32000, Israel

## Abstract

Reasoning about agent preferences on a set of alternatives, and the aggregation of such preferences into some social ranking is a fundamental issue in reasoning about uncertainty and multi-agent systems. When the set of agents and the set of alternatives coincide, we get the so-called reputation systems setting. Famous types of reputation systems include page ranking in the context of search engines and traders ranking in the context of e-commerce. In this paper we present the first axiomatic study of reputation systems. We present three basic postulates that the desired/aggregated social ranking should satisfy and prove an impossibility theorem showing that no appropriate social ranking, satisfying all requirements, exists. Then we show that by relaxing any of these requirements an appropriate social ranking can be found. We first study reputation systems with (only) positive feedbacks. This setting refers to systems where agents' votes are interpreted as indications for the importance of other agents, as is the case in page ranking. Following this, we discuss the case of negative feedbacks, a most common situation in e-commerce settings, where traders may complain about the behavior of others. Finally, we discuss the case where both positive and negative feedbacks are available.

## 1 Introduction

Reasoning about agent preferences on a set of alternative, and the aggregation of such preferences into some social ranking is a fundamental issue in reasoning about uncertainty and multi-agent systems. When the set of agents and the set of alternatives coincide, we get the so-called reputation systems setting.

In a reputation system agents are ranked based on other agents' reports on their quality of service, behavior, or importance. This topic has received a lot of attention in the recent multi-agent systems and e-commerce literature (see e.g. (Dellarocas 2003; Resnick *et al.* 2000)) . In particular, reputation systems have become a central ingredient of a variety of Internet sites, where perhaps the most famous examples are ebay's reputation system and Google's page rank.

Some reputation systems employ only positive feedback. These are in fact systems where agents point to the importance of other agents. For example, a pointer from page $p_1$ to page $p_2$ can be viewed as a positive feedback about the importance of page $p_2$. Here we get an interesting dynamics where we like to give more weight to the positive feedback provided by an important page than to the feedback provided by a less important page. In other reputation systems agents provide negative feedback with regard to the failure of participants in providing a particular service. A novel feature that distinguish such reputation systems for consumer to consumer (C2C) or peer-to-peer (P2P) settings from classical off-line reputation systems is the fact that individuals serve both as feedback providers, as well as the subject of such feedback. This again creates an interesting dynamics. In particular, if there is an agent who is reported to behave badly then one may like to consider less seriously his feedback with regard to bad behavior of others.

Reputation systems introduce a new social choice model. In the classical theory of social choice, as manifested by Arrow (Arrow 1963), a set of agents is called to rank a set of alternatives. Given the agents' input, i.e. the agents' individual rankings, a social ranking of the alternatives is generated. The theory studies desired properties of the aggregation of agents' rankings into a social ranking. In particular, Arrow's impossibility theorem shows that there is no aggregation rule that satisfies some minimal requirements, while by relaxing any of these requirements appropriate social aggregation rules can be defined. In the reputation systems setting the set of agents and the set of alternatives coincide. In addition, if e.g. agent $a$ provides positive (resp. negative) feedback on the importance of (resp. bad behavior of) agent $b$ then this may influence the credibility of a feedback by $b$ on the importance of (resp. bad behavior of) agent $c$; these indirect effects should be considered when we wish to aggregate the information provided by the agents into a social ranking.

In the social choice setting we discuss in this paper the input is a *reputation graph*. The reputation graph is a directed graph (with no self-loops) where nodes are associated with agents, and an edge pointing from node $a$ to node $b$ stands for a (positive or negative) feedback of $a$ on $b$. The output is a ranking (i.e. total pre-order) of the nodes of the graph. This ranking should satisfy some desired properties. In order to understand some of the fundamental differences between the classical social choice setting and the social choice setting induced by reputation systems consider the famous Independence of Irrelevant Alternatives (IIA) axiom, discussed in the classical theory of social choice. Roughly speaking this axiom says that by adding new alternatives and individual rankings of these alternatives with respect to existing alternatives, the social ranking of the previously existing alternatives should not change. This axiom is, and should be, false in any reasonable social choice setting induced by a reputation system; new alternatives that provide negative (resp. positive) feedback on an existing alternative might make it ranked lower (resp. higher) than another (previously existing) alternative.

Notice that a natural interpretation/application of the positive feedback setting is the ranking of Internet pages. In this case, if a page $p$ points to a set of pages $Q$ then in fact it declares that the set of important pages is $Q$. The problem of finding an appropriate social ranking in this case is in fact the problem of (global) page ranking. Particular approaches for obtaining a useful page ranking have been implemented by search engines such as Google. Our ideas may be used in order to tackle the axiomatic foundations of such rankings.

We consider three basic postulates for reputation systems. Below we consider a reputation system with only positive (resp. only negative) feedback.

**Generality:** The social ranking should be defined for any reputation graph.

**Transitivity:** If the set of agents that provide positive (resp. negative) feedback on agent $a$ is *more important* (resp. *more reliable*) than the set of agents that provide positive (resp. negative) feedback on agent $b$ then agent $a$ should be ranked higher (resp. lower) than agent $b$.

**Weak Monotonicity:** If the set of agents that provide positive (resp. negative) feedback on agent $a$ is not more important (resp. not more reliable) than the set of agents that provide positive (resp. negative) feedback on agent $b$, and $a$ is ranked socially higher (resp. socially lower) than $b$, then there should be at least one agent who provides positive (resp. negative) feedback on $a$ which is more important (resp. more reliable) than at least one agent who provides positive (resp. negative) feedback on $b$.

In the following sections we will formalize the above terms, and re-visit the postulates. We prove that there is no social choice rule that satisfies all the three postulates. However, we also prove that by relaxing any of these postulates we get that an appropriate social ranking exists. In particular, the relaxation of the weak monotonicity requirement leads to an efficient (and not immediate) algorithm for determining an appropriate social ranking. For the ease of exposition we will concentrate in Sections 2–4 on the case of reputation systems with only positive feedbacks. In section 5 we will return to the case of reputation systems with only negative feedbacks. In Section 6 we will discuss reputation systems where we have both positive and negative feedbacks. In Section 7 we discuss two natural refinements of the general setting, namely the restriction to strongly connected settings and weakening the monotonicity requirement, showing that these refinements will not change our basic results.

## 2 The Setting and Postulates

A (positive) reputation system setting is a directed graph $G = (V, E)$ (with no self-loops) where $V$ is a set of agents, and $E \subseteq V^2$ is a set of positive feedbacks. The set of agents who provide support for the importance of node $v$ is $\{u \in V : (u, v) \in E\}$. This set of "supporters" of node $v$ is denoted by $R(v)$. A social reputation rule $S$ associates with a reputation system setting, $G$, a social ranking, $S(G)$, of the corresponding nodes (in $V$). More formally, the social ranking is a function $S(G) : V \to N_{|V|}$ where $N_{|V|}$ stands for the integers between 1 and $|V|$. We denote the fact that $v_i$ is at least as preferable as $v_j$ in the social ranking, i.e. $S(G)(v_i) \leq S(G)(v_j)$, by $v_i \geq v_j$. We will use $v_i > v_j$ to denote the fact that $v_i \geq v_j$ but $\neg(v_j \geq v_i)$.

When we say that a node $v_i$ is ranked (socially) higher than a node $v_j$ we refer to the fact that $S(G)(v_i) < S(G)(v_j)$. In the context of positive reputation systems we will use the terms "as preferable as" and "as important as" interchangeably.

Given a reputation system setting $G = (V, E)$, and a social ranking $S(G)$, we will say that $R(v_i)$ is *more important* than $R(v_j)$, if there is a 1-1 mapping $f : R(v_j) \to R(v_i)$ such that for every $v \in R(v_j)$ we have that $v \leq f(v)$ and either $f$ is not onto or there exists at least one $v \in R(v_j)$ for which $v < f(v)$.

The intuition of the above is as follows. We wish to say that the more important are the agents that say that agent $a$ is important, then the more important $a$ is. In order to do this we need to refer to the importance of the supporters of different agents. This calls for a recursive approach for defining importance, which is in the spirit of approaches for defining importance in search engines. Assume we measured the importance of every agent in $R(v_i)$ and every agent in $R(v_j)$. If we can make a 1-1 mapping between $R(v_j)$ and $R(v_i)$ such that an agent in $R(v_j)$ will be always mapped to an agent who is at least as important as him, then $R(v_i)$ should be at least as important as $R(v_j)$. If, in addition, one of the agents in this mapping is mapped to an agent who is more important than he is, or the mapping does not cover all of the elements in $R(v_i)$, then we say that $R(v_i)$ is more important than $R(v_j)$. Similarly, we will say that $R(v_i)$ and $R(v_j)$ are equally important if there exists a 1-1 mapping as above, which is onto, and also $f(v)$ and $v$ have the same rank for every $v \in R(v_j)$.

Consider now the following central postulate:

- Transitivity [T]: Given a positive reputation system setting $G = (V, E)$ and a social ranking $S(G)$, then for every $v_i, v_j \in V$, if $R(v_i)$ is more important than $R(v_j)$ then $v_i > v_j$ (i.e. $v_i$ is more important than $v_j$).

The T postulate seems central to any rigorous theory of reputation systems. However, it leaves some cases quite unclear. For example, if there is one important agent who provides positive feedback on agent $a$, and two unimportant agents who provide positive feedback on agent $b$, then the T postulate may allow any ordering of $a$ and $b$. Some rationale for the social ordering in such cases is provided by the following general postulate:

- Weak Monotonicity [M]: Given a positive reputation system setting $G = (V, E)$, and a social ranking $S(G)$, then for every $v_i, v_j \in V$ we have that if $R(v_i)$ is not more important than $R(v_j)$ but $v_i > v_j$ then it must be the case that there exist $v_1 \in R(v_i)$ and $v_2 \in R(v_j)$ such that $v_1 > v_2$.

The M postulate says that you need to have a real reason for preferring one agent upon another if this is not implied by transitivity. In particular, if there is an important agent who says that $a$ is important, and two unimportant agents who claim that $b$ is important, then you can not prefer $b$ to $a$ and say that $b$ is more important (although you may rank $a$ and $b$ similarly).[1]

The last postulate is obvious:

- Generality [G]: A social reputation rule $S$ should associate with **every** reputation system setting $G$ an appropriate social ranking $S(G)$.

## 3 An Impossibility Result

In this section we prove the following general result:

**Theorem 1** *Given a positive reputation system setting, there is no social reputation rule that satisfies all of the postulates: $G, T,$ and $M$.*

**Proof (sketch):**

Consider agents $a, b, c, d$ where $a$ votes for $b$, $b$ votes for $c$, $c$ votes for $a$, and $d$ votes for $a$. Given transitivity agent $d$ should be ranked the lowest. Consider now agent $a$ and the way that it is ranked compared to agent $c$.

If agents $a$ and $c$ are ranked similarly then we get that $b$ is ranked higher than $a$ and $c$ (otherwise, we will contradict transitivity). But then (again by transitivity) $c$ should be ranked higher than $b$, which yields a contradiction.

If agent $a$ is ranked higher than agent $c$, then first notice it must be the case that agent $b$ is ranked lower than $a$: if agent $b$ is ranked higher than agent $a$ then agent $c$ will be ranked higher than $a$, which is a contradiction; if agents $a$ and $b$ are ranked similarly then this will imply that $b$ and $c$ should be ranked similarly, and therefore $a$ and $c$ should be ranked similarly, which also yields a contradiction. Therefore, when agent $a$ is ranked higher than agent $c$, then it must be the case that agent $b$ is ranked lower than agent $a$. Now however if agents $b$ and $c$ are ranked similarly then this will contradict transitivity ($b$ will have a stronger support than $c$); the same is true if agent $c$ is ranked higher than agent $b$. Therefore, we need to consider the case where agent $a$ is ranked higher than agent $c$, and agent $b$ is ranked in between. This however will contradict weak monotonicity: in this case all supporters of agent $a$ will have lower rank than the supporter of agent $b$.

---

[1] In Section 7 we show that replacing weak monotonicity by an extremely weak version of it will not change our results.

If agent $a$ is ranked lower than agent $c$ then this can only be if agent $b$ is ranked at least as high as agent $c$ (since otherwise the support for $a$ will be stronger than the support for $c$, which will contradict transitivity). However, agent $b$ can not be ranked higher agent $c$ (since he has only agent $a$ in his support); so, we remain with the case where agents $b$ and $c$ are ranked similarly, but higher than $a$. This will however contradict the fact that $c$ is preferable to $a$ (since now $a$ has a stronger support).

## 4 A Possibility result

One may ask whether all the previous conditions are necessary for obtaining the previous impossibility result. Moreover, one may say that in certain cases some of the postulates may be relaxed. For example, one may argue that if only $a$ provides positive feedback on $b$, then even if $a$ is the most important agent in the system then if there are many agents (who are less important than $a$) who provide positive feedback on $c$, then $c$ should be ranked higher than $a$; in this case we may wish to relax weak monotonicity.

It is trivial to see that when relaxing $G$ (generality) an appropriate social ranking exists. It is also quite immediate to see that by relaxing $T$ (transitivity) we always get the desired ranking. In this case if all agents are ranked similarly then we get an appropriate social ranking. It should not be a surprise that the question of whether relaxing $M$ is sufficient for obtaining the desired ranking is the most subtle. We can show:

**Theorem 2** *We can always satisfy any pair of the postulates: $G, T$, and $M$.*

**Proof (sketch):**

We will show that we can relax $M$. The other cases are rather straightforward as mentioned before. We consider the following algorithm for ranking the nodes of a reputation graph $G = (V, E)$. Let $v_i^l \geq v_j^l$ denote the fact that the social rank of $v_i$ is greater than or equal to the social rank of $v_j$ in iteration $l$ of the algorithm.

1. Let k:=0;

2. Let $v_i^0 \geq v_j^0$ iff $|R(v_i)| \geq |R(v_j)|$

3. If there exist $v_i$ and $v_j$ such that $v_i^k = v_j^k$ but $R(v_i)$ is more important than $R(v_j)$ according to the $v_l^k (1 \leq l \leq |V|)$, and there is no $v_s$ such that $v_s^k = v_i^k$ and $R(v_s)$ is more important than $R(v_i)$ then do(if there is more than one $v_i$ and $v_j$ as above, then choose one of them) :

    (a) Let $v_i^{k+1} > v_l^{k+1}$ for every $v_l$ such that $v_l^k = v_i^k$ and $R(v_l^k)$ and $R(v_i^k)$ are not equally important.
    (b) Let $v_i^{k+1} = v_l^{k+1}$ for every $v_l$ such that $v_l^k = v_i^k$ and $R(v_l^k)$ and $R(v_i^k)$ are equally important.
    (c) Let $v_i^{k+1} < v_l^{k+1}$ (resp. $v_i^{k+1} > v_l^{k+1}$) iff $v_i^k < v_l^k$ (resp. $v_i^k > v_l^k$) for every $v_l$ such that $v_l^k \neq v_i^k$.
    (d) Let the relation between $v_q^{k+1}$ and $v_r^{k+1}$, for every $v_q \neq v_i$ and $v_r \neq v_i$ be as the relation between $v_q^k$ and $v_r^k$, unless $v_r^k = v_i^k$ and $R(v_r^k)$ and $R(v_i^k)$ are equally important, or $v_s^k = v_i^k$ and $R(v_s^k)$ and $R(v_i^k)$ are equally important.

4. If there is no such pair then halt, and output the current ranking as the chosen one.

5. Let k:=k+1, and go to 3.

Notice that the ranking in stage $k + 1$ is a refinement of the ranking in stage $k$, where all we may do is make $i$ preferable to $j$ where they were ranked similarly before. If $R(v_i)$ is more important than $R(v_j)$ in iteration $k$, then it can not become less important than $R(v_j)$ in the future . This is implied by the definition of "more important". In order that $R(v_i)$ will be more important than $R(v_j)$ in iteration $k$ then we need to show a 1-1 correspondence between $R(v_j)$ and $R(v_i)$ that will map the elements in $R(v_j)$ to elements with equal or higher ranks in $R(v_i)$, where we have one strict preference or the mapping is not onto. When we refine the rankings in future steps, some of the equalities in the 1-1 mapping may become inequalities but still $R(v_j)$ can not become more important than $R(v_i)$. In addition, notice that when $v_i$ is selected in iteration $k$, then each agent $v_j$ with $R(v_j)$ of similar importance to $R(v_i)$ at this stage is ranked similarly to $v_i$. As a result, by iteratively applying the transitivity requirement we can not contradict future changes in the coming iterations.

Notice that since any iteration is just a refinement that makes the ranking more strict, then this procedure will be over after $O(|V|)$ iterations. The final output will be consistent with transitivity by the arguments above, and by the fact that if there exist $v_i$ and $v_j$ whose relative ranking should be refined to satisfy transitivity then the procedure will not stop and will make further refinements.

Further formal details are omitted due to lack of space and will appear in the full paper. ∎

Intuitively, the algorithm which allows satisfying transitivity for arbitrary graphs makes use of the following idea. We first sort agents based on their in-degree, the higher is the in-degree of a node the higher is its rank. Then, we recursively refine the rankings, i.e. make indifference into strict preference, while keeping previously existing strict preferences, by applying transitivity. The fact we use only such refinements will prevent us from entering into cycles. Assume that agent $i$ has been ranked higher than $j$ in the above process. If the size of the support set of $i$ is greater than the size of the support set of $j$ then no further refinements will lead to a situation where the support set of $j$ is more important than the support set of $i$. Otherwise, if a refinement made $i$ preferable to $j$ then consider the sorted list (from low to high) of the ranks of the supporters of $i$ and $j$, respectively. Consider the first agent, $v$, in the (sorted) support set of agent $i$ such that the corresponding agent, $v'$, in the (sorted) support set of agent $j$ is ranked lower than it. Such an agent must exist, based on our construction. In this case, no further refinements will be able to make $v$ ranked lower or equal to $v'$ (since, again, refinements do not change previous strict preferences). This kind of reasoning is the basic intuition behind the fact we get no cycles. Other parts of the algorithm are needed for full correctness (e.g. the fact we need to take care that agents with support sets of the same strength will be ranked similarly etc.).

## 5   Reputation systems with negative feedbacks

In order to introduce reputation systems with negative feedbacks, we need to understand the subtle difference between the cases of positive and negative feedbacks. Consider a system with three agents $a, b, c$, and a reputation graph where $a$ points to $b$ and $b$ points to $c$. When the pointers refer to positive feedbacks then it is obvious that $c$ is the most important, while $b$ is more important than $a$. This is in fact a simple instance of the transitivity requirement we had before. On the other hand, the logic in the case where the pointers refer to negative feedbacks may (and should) be quite different. Agent $a$ is the only one that no one provides negative feedback about him, and therefore he should be ranked first. Now we need to compare $b$ and $c$, but since $a$ is the most reliable agent (i.e. ranked first) then agent $b$ should be ranked the lowest and agent $c$ should be ranked in the middle. This means that the transitive effects here are of very different nature than in the case of positive feedbacks. As a result, although our definitions, results, and algorithm for the positive and negative reputation settings, may seem relatively similar to one another, the "local" differences between these cases have significant effects, and should not be neglected.

Now we have a (negative) reputation system setting, which is a directed graph $G = (V, E)$ (with no self-loops) where $V$ is a set of agents, and $E \subseteq V^2$ is a set of negative feedbacks. For every $v$ let $R(v)$ be as before, now standing for the set of agents who provide negative feedback on $v$. Given a negative reputation system setting $G = (V, E)$, and a social ranking $S(G)$, we will define when $R(v_i)$ is *more reliable* than $R(v_j)$ similarly to the way we defined "more important" before (all changes until this point are only in the interpretation of the edges).

Now we wish to say that the more reliable are the agents that say that $a$ is unreliable, the more unreliable $a$ is. Similarly, the more unreliable these agents are, the better $a$'s reputation is. Of course, if no one provides negative feedback with regard to $a$, then he is indeed very reliable. The above discussion leads to a revised transitivity postulate. Similarly, it leads to a revised weak monotonicity postulate. These are stated below:

- B-Transitivity [BT]: Given a negative reputation system setting $G = (V, E)$, and a social ranking $S(G)$, then for every $v_i, v_j \in V$ we have that if $R(v_i)$ is more reliable than $R(v_j)$ then $v_i < v_j$ (i.e. $v_i$ is *less* reliable than $v_j$).

- B-Weak-Monotonicity [BM]: Given a negative reputation system setting $G = (V, E)$, and a social ranking $S(G)$, then for every $v_i, v_j \in V$ we have that if $R(v_i)$ is not more reliable than $R(v_j)$ but $v_i < v_j$ then it must be the case that there exist $v_1 \in R(v_i)$ and $v_2 \in R(v_j)$ such that $v_1 > v_2$.

The Generality postulate will be defined as before, but now it is applied to negative reputation system settings.

### 5.1 An (im)possibility result – the negative feedbacks setting

In this section we prove the following general result:

**Theorem 3** *Given a negative reputation system setting, there is no social reputation rule that satisfies all of the postulates: $G, BT$, and $BM$. By relaxing any one of these postulates we get an appropriate social ranking.*

**Proof (sketch):**

First we show the impossibility result.

Consider the graph $G(V, E)$ with nodes $a, b, c, d$, and the following edges (reports): $a$ points to $b$, $b$ points to $c$, $c$ points to $d$, $d$ points to $a$, $a$ points to $c$.

Consider the relative ranking of $a$, $b$, and $d$.

If $b = d$ then transitivity implies that it must be that $a = c$ (since $a$ and $c$ are the only ones that point to $b$ and $d$, respectively). However, by transitivity it also must be the case that $a > c$ (since $b = d$ and in addition $a$ points to $c$), and therefore we get a contradiction.

If $a = d$ then transitivity implies $a = d = b$, and that $c < a = b = d$. This however should imply that $d$ (who is pointer to only by $c$) should be ranked higher than $a$ (who is reported only by $d$), which yields a contradiction.

If $a = b$ then transitivity implies that $c < b = a$, and therefore (again by transitivity) that $d > b = a$. Now however, $a$ and $b$ are ranked similarly, while $a$ has a more reliable support, which yields a contradiction.

The above cases show that $a, b, d$ should appear in three different ranks, and we consider below all 6 possibilities.

Assume that $d > a > b$ then this immediately contradicts transitivity, since $d$ is the only one that points to $a$ and $a$ is the only one who points to $b$, while $a$ is ranked higher than $b$.

Assume that $d > b > a$. If $c \geq a$ then it should be that $b \geq d$, and therefore it must be that $c < a$. But now, since $d$ is ranked higher than both $a$ and $b$, the fact $c < a$ contradicts weak monotonicity.

Assume that $a > d > b$. Monotonicity implies that $c < b$. But this contradicts the fact that $a > d$, since $c$ is ranked lower than $d$, and $c$ (resp. $d$) is the only one that points to $d$ (resp. $a$).

Assume that $a > b > d$. Then transitivity implies that $c < b$. This contradicts however the fact that $b > d$, since $c$ is the only one that points to $d$, and is ranked lower than $a$, who is the only one who points to $b$.

Assume that $b > d > a$. Then, transitivity implies that $c < a$. However, now $c$, who is the only pointer to $d$, where $d < b$, is ranked lower than $a$, who is the only pointer to $b$, which contradicts transitivity.

Assume that $b > a > d$. This immediately contradicts transitivity, since both $a$ and $b$ have only one agent ($d$ and $a$, resp.) that point to them.

To prove possibility when relaxing any of the postulates, we observe again that relaxing the weak monotonicity requirement is the most subtle. We use a slightly revised version of the algorithm for the positive feedback case.

The reader should notice that although the axiomatization for the cases of positive and negative feedbacks are similar, the connections between these cases are not immediate. In particular, we can show that there are graphs for which, when considering "positive feedback", no appropriate social ranking exists (if we wish to satisfy all requirements), while when we consider the complement graph and interpret it as "negative feedbacks" then an appropriate social ranking (which satisfies all related requirements) does exist. In the complement graph an agent will provide negative feedback on the set of agents on which he did not provide positive feedback in the original graph (excluding itself; we assume both graphs do not have self-loops).

To see the above consider the following setting. Agents: $a, b, c, d$. Pointers: $a$ points to $b$, $b$ points to $c$, $c$ points to $d$, $d$ points to $b$.

One can check that in this case we do not have any social ranking that satisfies transitivity and weak monotonicity when one interprets the pointers as "positive feedback". However, in the reputation system with "negative reports", which refers to the complement graph, we get that $d > c > b > a$ is an appropriate social ranking.

## 6 Reputation systems with both positive and negative feedbacks

Our results can be extended to the case where we have two types of pointers: "positive feedback" and "negative feedback". In this case we have two types of incoming edges to node $v_i$: the negative reports with regard to $v_i - R_b(v_i)$, and the positive reports with regard to $v_i - R_g(v_i)$.

Transitivity will be defined as follows: we will say that $R(v_i)$ is socially stronger than $R(v_j)$ if $R_b(v_i)$ is either less reliable or have the same reliability as $R_b(v_j)$ and $R_g(v_i)$ is either more important or as important as $R_g(v_j)$, with at least one strict inequality. We now require that if $R(v_i)$ is socially stronger than $R(v_j)$ then $v_i$ should be ranked higher than $v_j$. Weak monotonicity will require that at least weak monotonicity with regard to $R_b$ or with regard to $R_g$ should hold, i.e. if $R(v_i)$ is not socially stronger than $R(v_j)$ but $v_i$ is ranked higher than $v_j$ then either there are $v_1 \in R_g(v_i), v_1' \in R_g(v_j)$ such that $v_1 > v_1'$ or there exist $v_2 \in R_b(v_i), v_2' \in R_b(v_j)$ such that $v_2 < v_2'$.

We will refer to a reputation system setting with both negative and positive feedbacks as: complete reputation system setting. We will denote the corresponding transitivity, weak monotonicity, and generality postulates in this case by $T_c, M_c, G_c$, respectively.

**Theorem 4** *It is impossible to satisfy all 3 postulates: $T_c, M_c, G_c$ for complete reputation systems settings with both positive and negative feedbacks. Relaxing any of these postulates will allow generating appropriate social rankings.*

**Proof (sketch)**:

The impossibility result is implied by the previous impossibility results. In order to show that we can get $G_c$ and $T_c$ (without $M_c$) we use similar ideas to the algorithm presented for positive feedbacks, employing the new definition of "socially stronger".

## 7 Discussion: strongly connected systems and very weak monotonicity

We have seen that if we adopt all our three basic postulates then an appropriate impossibility result is obtained, while any pair of these postulates can be satisfied. We now consider the idea of weakening the postulates without completely removing any of them. In particular, one may wish to consider less general settings where not every reputation graph is possible, and/or some weakening of the weak monotonicity postulate. We will concentrate below on the case of positive reputation settings, although our findings do hold for the other settings discussed in this paper as well.

Two major refinements suggested to us by researchers interested in the practice of page ranking are the following:

1. It may be useful to consider only strongly connected reputation graphs, where there is a directed path from node $v$ to node $t$ for any pair of nodes $v$ and $t \neq v$. This is due to the fact that these are graphs where page ranking tends to "behave well". We will call the related generality assumption: weak generality [WG].

2. It may be useful to restrict the weak monotonicity assumption by requiring that it will hold only for nodes $v_i$ and $v_j$ which have supports of more or less the same size. This is done in order to allow a node $v_i$ to be socially ranked higher than $v_j$, when this is not implied by transitivity and there is no element in $R(v_i)$ which is ranked higher than some element in $R(v_j)$, if the size of the support of $v_i$ is much larger than the support of $v_j$. A very weak (and therefore satisfactory for the result below) version of this, that we call very weak monotonicity [VWM] is the following: Given a positive reputation system setting $G = (V, E)$, and a social ranking $S(G)$, then for every $v_i, v_j \in V$ where $v_i > v_j$, if $R(v_i)$ is not more important than $R(v_j)$ and $|R(v_i)| \leq |R(v_j)| + 1$, it must be the case that there exist $v_1 \in R(v_i)$ and $v_2 \in R(v_j)$ such that $v_1 > v_2$.

We can show:

**Theorem 5** *Given a positive reputation system setting, there is no social ranking that satisfies all of the postulates: $WG, T,$ and $VWM$.*

The proof of Theorem 5 uses a similar construction to the one used in Theorem 3, which is based on a strongly connected setting. The fact we require very weak monotonicity does not prevent us from reaching impossibility since the maximal difference in the in-degree of nodes in the above construction is at most one.

Similar results can be presented for reputation systems with negative feedbacks, and for settings where we have both positive and negative feedbacks. The above result suggests that small "natural" refinements of the basic postulates do not suffice for obtaining a possibility result, while such a result is obtained by dropping any single postulate.

## 8 Conclusion

Reasoning about preferences and preference aggregation is a fundamental task in reasoning about uncertainty and multi-agent systems (see e.g. (Boutilier *et al.* 2003; Conitzer & Sandholm 2002; Yager 1991; LaMura & Shoham 1998). When the set of agents and the set of alternatives coincide, preferences among alternatives can be viewed as feedback by the agents on the importance or the reliability of other agents. This brings us to the study of reputation systems. Reputation systems are fundamental ingredients of some of the most famous tools/techniques in the Internet (e.g. Google's page rank and ebay's reputation systems, among many others). Moreover, the task of building successful and effective on-line trading environments has become a central challenge to the AI community (Boutilier, Shoham, & Wellman 1997; Monderer, Tennenholtz, & Varian 2000; Sandholm 2003). Feedback mechanisms are believed to be fundamental for the establishment of such environments. Although reputation has always been a major issue in economics (see e.g. (Kreps & Wilson 1982; Milgrom & Roberts 1982)), reputation systems have become so central recently due to the fact that some of the most influential and powerful Internet sites and companies have put reputation systems in the core of their business.

Our aim in this paper was to to treat reputation systems from an axiomatic perspective. The classical theory of social choice lay the foundations to large part of the rigorous work on multi-agent systems. Indeed, the most classical results in the theory of mechanism design, such as the Gibbard-Satterthwaite (Gibbard 1973; Satterthwaite 1975) are applications of the theory of social choice. Moreover, previous work in AI has employed the theory of social choice for obtaining foundations for reasoning tasks (Doyle & Wellman 1989) and multi-agent coordination (Kfir-Dahav & Tennenholtz 1996). It is however interesting to note that reputation systems suggest a novel and new type of theory of social choice. We see this point as especially attractive, and as a main reason for concentrating on the study of the axiomatic foundations of reputation systems.

In this paper we attempted to capture basic ingredients of what we believe is a new axiomatic theory of social choice. Of course, much is left to be done, and we hope that others will join us in addressing the challenge of establishing a rigorous axiomatic approach to reputation systems.